\documentclass[aps,11pt,pra,showpacs,tightenlines]{revtex4}
\usepackage{graphicx}

\begin{document}

\title{Reflection from a moving mirror - a simple derivation \\using the photon model of light}

\author{Aleksandar Gjurchinovski}

\email{agjurcin@pmf.ukim.mk}

\affiliation{Institute of Physics, Faculty of Natural Sciences 
and Mathematics, Sts.\ Cyril and Methodius University,
P.\ O.\ Box 162, 1000 Skopje, Macedonia}

\begin{abstract}
We present an elementary analysis of the effects on light reflected from a uniformly
moving mirror by using the photon picture of light and the conservation 
laws for energy and momentum of the system photon-mirror.
Such a dynamical approach to the problem seems very suitable for introductory physics courses, not requiring any previous knowledge in wave optics or special relativity.
\end{abstract}

\pacs{42.15.-i, 03.30.+p}

\date{June 28, 2012}

\maketitle

The reflection of light from a moving mirror has been studied 
in many different contexts during the last century \cite{shankland,abraham,
einstein,majorana,kennard,kennedy,yeh,fulling,davies,anglada,sutanto}. 
In contrast to the usual problem of reflection from a stationary mirror, 
the situation with the moving mirror give rise to several important 
deviations from the classical case \cite{bolotovskii}. 
In fact, a very interesting phenomenon in quantum field theory that involves moving mirrors, 
which has been only recently observed experimentally, is the so-called dynamical Casimir effect, 
in which a mirror undergoing relativistic motion converts virtual photons into directly 
observable real photons \cite{wilson}. 

In this paper we will concentrate on two effects regarding reflection from a uniformly moving mirror.
The first is that the angle of incidence is no longer equal to the angle of reflection 
\cite{einstein,gjurchinovski1,gjurchinovski2,gjurchinovski3}. The second is the so-called
Doppler effect upon reflection from the moving mirror, or a change in the frequency of the reflected light with respect to the 
incident one \cite{einstein,gjurchinovski4}. 
In our approach we will use the photon picture of light and investigate the collision of the photon
with the moving mirror using the conservation laws governing elastic collisions \cite{wichman}.

Consider a single photon of energy $\hbar\omega$ and a corresponding momentum of magnitude $\hbar\omega/c$,
upon oblique incidence on a perfectly-reflecting vertical plane mirror moving at a constant speed $v$ to the
right (see Fig. 1).
\begin{figure}
\includegraphics[width=0.5\columnwidth,height=!]{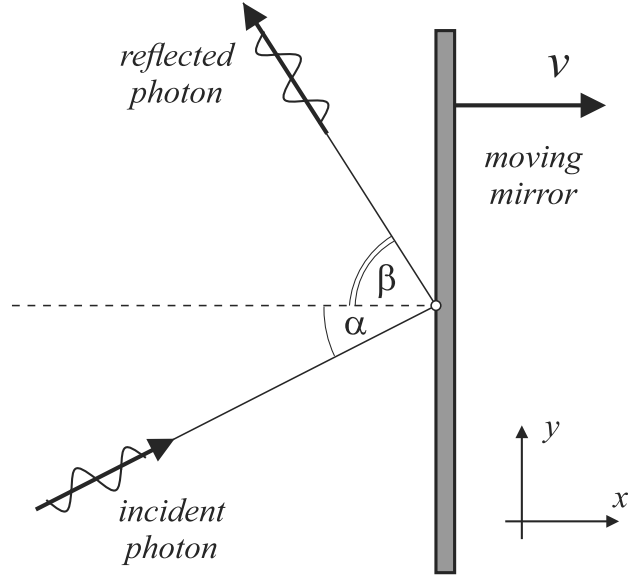}
\caption{Reflection of a photon from a massive plane mirror in uniform rectilinear motion.}
\end{figure} 
According to the quantum picture of the process of reflection, the photon will be 
absorbed and reemitted by the target atoms from the surface of the mirror \cite{wichman}. The motion of the mirror
will cause loss or gain of momentum (and, therefore, energy) of the photon upon reflection, depending whether
the mirror is moving along the positive or the negative direction of the $x$-axis. 
In the following, we shall assume that the motion of the mirror is non-relativistic ($v<<c$), and that the reflection 
of the photon at its surface is perfectly elastic. The latter assertion allows us to treat the
photon as a spherical particle that collides and bounces-off a frictionless plane surface.

As a consequence of the interaction between the photon and the moving mirror, there will be a shift in the frequency of the
photon after the reflection, and the incident angle $\alpha$ and the reflected angle $\beta$ will not be equal. 
Let $\hbar\omega'$ and $\hbar\omega'/c$ denote the respective energy and momentum magnitude of the reflected photon.
Applying the conditions for conservation of momentum along $x$ and $y$ axes, we get:
\begin{eqnarray}
\frac{\hbar\omega}{c}\cos\alpha+Mv&=&-\frac{\hbar\omega'}{c}\cos\beta+M(v+\Delta v_x),\\
\frac{\hbar\omega}{c}\sin\alpha&=&\frac{\hbar\omega'}{c}\sin\beta,
\end{eqnarray}
where $M$ is the mass of the mirror.
We have taken into account that the velocity of the mirror is changed after the reflection
by an amount $\Delta v_x$ in the $x$ direction. There is no changing of the mirror's velocity 
along the direction of its surface (in the $y$ direction) under the assumption that the collision 
is perfectly elastic. 
We further apply the law of conservation of energy:
\begin{equation}
\hbar\omega+\frac{Mv^2}{2}=\hbar\omega'+\frac{M(v+\Delta v_x)^2}{2},
\end{equation}
which can be easily transformed into:
\begin{equation}
\hbar\omega=\hbar\omega'+v(M\Delta v_x)+\frac{(M\Delta v_x)^2}{2M},
\end{equation}
From Eq. (1) we have:
\begin{equation}
M\Delta v_x=\frac{\hbar}{c}\left(\omega\cos\alpha+\omega'\cos\beta\right),\\
\end{equation}
which upon substitution in Eq. (4) yields:
\begin{equation}
\hbar\omega=\hbar\omega'+\frac{\hbar v}{c}\left(\omega\cos\alpha+\omega'\cos\beta\right)
+\frac{\hbar^2}{2Mc^2}\left(\omega\cos\alpha+\omega'\cos\beta\right)^2.
\end{equation}
In the limiting case of an infinitely heavy mirror ($M\rightarrow\infty$)
the last term in the right-hand side of Eq. (6) can be neglected, and we obtain:
\begin{equation}
\omega'=\omega\frac{1-v\cos\alpha/c}{1+v\cos\beta/c}.
\end{equation}
Consequently, Eq. (2) can be recast into:
\begin{equation}
\sin\beta\left(1-\frac{v}{c}\cos\alpha\right)=\sin\alpha\left(1+\frac{v}{c}\cos\beta\right),
\end{equation}
which can be solved for the reflection angle $\beta$ in terms of the angle of incidence $\alpha$ and 
the velocity $v$ of the moving mirror to give:
\begin{equation}
\cos\beta=\frac{-2v/c+(1+v^2/c^2)\cos\alpha}{1-2v\cos\alpha/c+v^2/c^2}.
\end{equation}
Furthermore, a substitution of Eq. (9) into Eq. (7) yields the relation between the frequencies
of the photon before and after the reflection:
\begin{equation}
\omega'=\omega\frac{1-2v\cos\alpha/c+v^2/c^2}{1-v^2/c^2}.
\end{equation}

The reflection laws (9) and (10) coincide with the formulas obtained by using 
relativistic electrodynamics, the principles of wave optics, or directly from the postulates of special relativity 
\cite{einstein,bolotovskii,stephani,pauli,synge,yeh,gjurchinovski1,gjurchinovski2,gjurchinovski3,gjurchinovski4}.  
The interesting point is that in deriving the same equations, we only use the relations for the energy 
and momentum of light in the sense of the photon model, which were apparently discovered and developed 
independently of special relativity \cite{duck}. 

Another curious point is that the resulting formulas (9) and (10) of our essentially non-relativistic approach 
are also valid in the case when the mirror is moving at relativistic speeds. The student is encouraged to prove this 
result rigorously by repeating the above derivation using the relativistic forms of conservation laws for energy and momentum.

\end{document}